\documentclass{kluwer} 
\usepackage{graphicx}
\usepackage{epigraph}
\newcommand {\eqrefa}[1]{(\ref {#1})}
\usepackage{url}
\begin{document}

\begin{opening}

\title{Time Travel and the Reality of Spontaneity\thanks{  
A preliminary version of this paper was presented at a Seminar on 
`Reality in Physics and Philosophy,' S. N. Bose National Centre for 
the Basic Sciences, Calcutta, 24-25 Feb 1996.}}

\author{C. K. Raju\thanks{Present address: Centre for Computer Science, MCRP University, 222 M. P. Nagar Zone 1, Bhopal 462 011.}   }
\runningauthor{C.\ K.\ Raju}
\runningtitle{Time Travel and Spontaneity}
\institute{NISTADS, Pusa, New Delhi 110 012\\ email: \url{c_k_raju@vsnl.net} \vspace{1 in}}
\begin {abstract}
Contrary to the informed consensus, time 
travel implies spontaneity (as distinct from chance) so that time 
travel can only be of the second kind. 
\vspace{1in}
\end{abstract} 


\end{opening}

\maketitle 

\pagebreak

\epigraph{No patient who recovers without a physician can logically 
attribute his recovery to spontaneity. Indeed, under a close examination 
spontaneity disappears. For everything that occurs will be found to 
do so through something, and this `through something' shows that spontaneity 
is a mere name and has no reality. Medicine, however, because it acts 
`through something' and because its results may be forecasted, has 
reality.\footnote{Hippocrates, \textit{ The Art}, cited in 
S. Sambursky, \textit{Physics of the Stoics} (Routledge 
and Keegan Paul, London, 1987) pp. 51--52.} }

\pagebreak

\section{INTRODUCTION}

\noindent
In  this note I consider time travel both as  a  real 
physical  possibility  and as a means  of  re-examining fundamental 
assumptions about time. Though stemming from a new mathematical model 
of the evolutionary equations of physics, the arguments in this note 
are robust enough to be stated with the technicalities only in the 
background. Such a style of exposition also seems desirable in view 
of the widespread interest in time-travel. 

\subsection{Background}

\noindent
Thorne and his consortium have proposed$^1$ time 
machines based on `wormhole'  solutions,  exploiting the  fact  that  the 
Hilbert-Einstein equations  are  silent   about the (algebraic)  topology 
of  spacetime. While  the  `wormhole'   solutions  involve  `exotic   matter'---matter 
with negative mass  and  positively  amusing  properties$^2$---Gott$^3$  has 
shown that closed timelike curves  (CTCs) may also arise with cosmic 
strings.  On the  other hand, Hawking$^4$ has argued that there 
is  excellent empirical  evidence for chronology protection since  we 
have  not been invaded by hordes of tourists  from  the future.  

\subsection{Two Kinds of Time-Travel}

\noindent
For  the  purposes  of this note it helps  to  make  an 
informal distinction between two types of time  travel: (i)  with 
time-machines and (ii) without  machines.  An example  of the second 
kind of time travel is  transfer of  information  using  a  retarded  interaction  going 
forward  in time and an advanced interaction  returning backward in 
time.$^5$ Access to advanced interactions$^6$  is possible 
under the hypothesis of a microphysical tilt in the arrow of time.$^7$ 
Strictly speaking, a `tilt' does not involve any new hypothesis; the 
usual hypothesis of `causality'  is rejected,  so  that the evolution  of  a  many-particle 
system  is governed by a different category  of  (mixed-type functional 
differential) equations of motion. 

Time  travel  of  the  second  kind  contemplates  only transfer  of  information  without  involving  physical 
transport  of the traveler's body.  Nevertheless,  some (diminished)  kind  of  intervention  in  the  past  is 
possible,  in  principle, because  information  may  be transferred   from  present  to  past  using   advanced 
interactions,  though  the bandwidth is  a  very  small fraction  of 
the bandwidth for information transfer  to the future using retarded 
interactions. 

\subsection{Aim}

\noindent
The  aim  of  this note is to stand  on  its  head  the 
standard conclusion derived from the paradoxes of  time travel,$^8$ 
especially for the case of time travel without machines. 

\section{THE PARADOXES OF TIME TRAVEL}

\subsection{The Grandfather Paradox}

\noindent
The  grandfather  paradox is well-known:$^9$  Tim  travels 
into  the past to kill his grandfather when  Grandfather  was  yet 
a boy; but that would mean that Tim could not  have  been born and 
so could not have killed Grandfather. The generally accepted conclusion 
is as  follows. Since  Tim did not kill Grandfather in  the  `original' 
1921,  consistency  demands that neither does  he  kill Grandfather  in 
the `new' 1921. The time traveler  must be  prepared for unexpected 
constraints; Tim must  fail in  the  enterprise  of killing  Grandfather  for  some 
commonplace reason. Perhaps some noise distracts him at the last moment, 
perhaps he misses despite much  target practice,  perhaps even Tim 
killed Grandfather only  to discover   his  true  antecedents!  As  summarised   by 
Woodward,$^{10}$ `Time travel makes ``free will" irrelevant'. 

\subsection{Mundane Time}

\begin{figure}[h] 
\centerline {
\includegraphics 
		[width = 18pc] {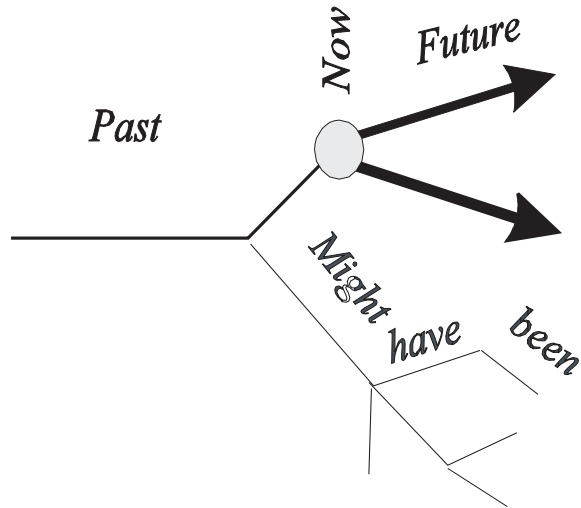}
}
\caption {\textbf{Mundane time}: In everyday life, one philosophizes about the past but 
agonizes about the future on the belief  that one's actions partly 
decide the future, but leave the past unaffected.  
}
\label {mundane}
\end{figure}

\begin{figure}[h]
\centerline{
\includegraphics
	[width=18pc]{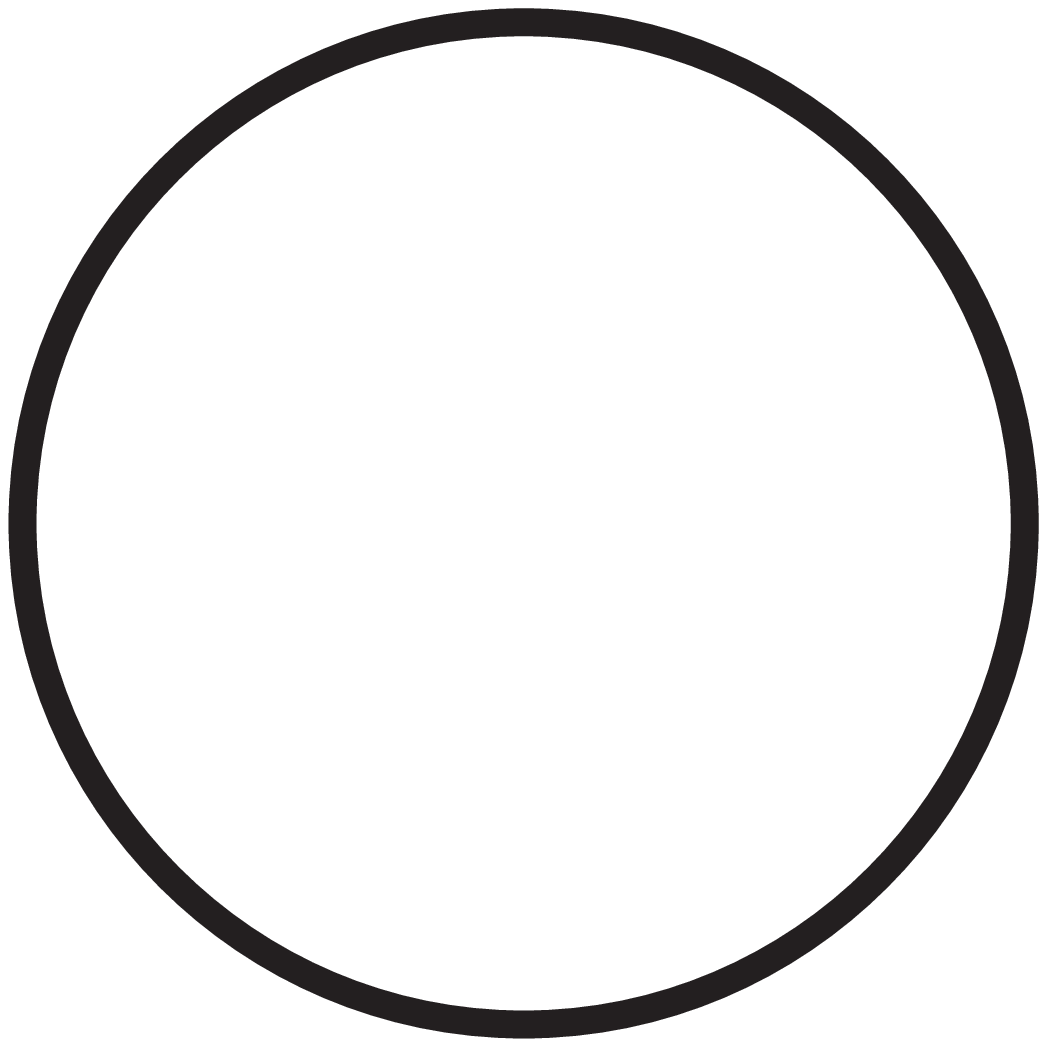}}
\caption{\textbf{Supercyclic time}: All instants of time are arranged in a closed cycle, 
so that any instant `precedes' any other. Such a situation cannot 
be readily described with the binary earlier-later relation implicit 
in the tense-structure of Indo-European languages.}
\label{circ}

\end{figure}

\noindent
One  could  elaborate the paradox as  follows.  Mundane 
time  has a structure$^{11}$ which is past linear and  future branching 
(Fig. \ref{mundane}). If one bends it around in a  circle and  joins future to 
past then either future  branching or past linearity must fail, so 
that one obtains the supercyclic time of Fig. \ref{circ}. 

\subsection{Popper's Record Postulate}

\noindent
Why not give up past linearity? This could be  problematic,  since 
the significance of  experimental  records would  then diminish, for 
an experimental record could not, then, claim to represent \textit{the} 
past.  Popper$^{12}$ proposed a record  postulate,  the  `principle 
of the  unbroken  connection  of world  lines' which he formulated 
in operational  terms as follows. 

\begin{quote}
Any `observer' (local material system) can begin,  at any 
instant, a record (causal  trace);  make successive  entries into 
that record; and  arrange  for the  preservation of the record for 
any desired  finite period  of time. (By `can' the following is meant;  the 
theoretical  possibility  of  any  world-line,  to   be considered 
as consistent with the laws of nature,  must not  entail  the impossibility 
of  the  operations  described in the above principle.) 
\end{quote}

World lines closed in time now lead to a contradiction, since, for 
consistency, the closed world line `must  be infinitely and absolutely 
repetitive,' and hence `would entail  periodic destruction of every  single  record,' 
since otherwise the record `would not be fully  repetitive  but  would 
constantly enrich  itself  upon  every renewal of the closed journey.'

\subsection{The Chronology Condition}

\noindent
Appeal to the future branching alone is also  adequate. 
`The  same result may be obtained, even less  \textit{ad  hoc},' continues 
Popper, `by adopting a ``principle of indeterminism";  this  too  would  automatically  exclude  all 
cosmological solutions permitting closed  world-lines.' Hawking and 
Ellis$^{13}$ similarly argue that future  branching  cannot  be  lightly 
rejected, since  `all  of  our philosophy  of science is based on 
the assumption  that one is free to perform any experiment.' Hence, 
they are `much  more ready to believe' their  \textit{chronology  condition},  viz. 
that there are no CTCs.  (Hawking's  latest position, marks a retreat 
from postulate to conjecture, and  adds  the bit about making the 
universe  safe  for historians.) 

\subsection{The Paradoxes Re-examined}

\noindent
In  brief, the informed consensus favours the standard 
conclusion$^{14}$ that time travel  is antithetical  to  spontaneity 
or `free  will'.  I  will argue that the exact opposite is true.  

Let  us re-examine the grandfather paradox, for two  of its  key features 
seem to have gone unnoticed. We~need to shift our attention from 
the death of Grandfather to the  birth of Tim, that is to the first  appearance  of 
Tim  in this world. Let us suppose that Tim's `birth' (i.e. his chronologically 
earliest appearance in the world) was earlier than his biological 
birth from his mother's womb. Let us further suppose that the event 
of Tim's `birth'  did  not  go unobserved. Say, Tim's house had earlier 
been  occupied by an eccentric scientist, who had called another half-a-dozen  scientists 
for tea. Tim, being a tyro at  time travel,  appeared bang in the 
midst of this tea  party. The  scientists,  true  to  their  profession,   merely 
observed  and  theorised: they did not  hop  around  or interfere  in 
what they took to be a  demonstration  to challenge  their  theoretical  capabilities,  specially 
arranged  by their eccentric host (who had  disappeared into  the  kitchen).  Naturally, 
they  were  all  blase enough to regard it as a magic trick in bad 
taste. (Tim materialised with one foot on a saucer, and spilled tea 
on  a  guest.)  

But we know better. We  know  that, however hard they might have tried,  the 
scientists could not have found an explanation for  the fact  which  was  presented 
to them  on  a  platter---no causal explanation that is. We know 
that Tim's  appearance at the eccentric tea party really had nothing 
to do  with  anything  prior  to the  tea  party;  it  was causally  inexplicable,  spontaneous, 
so  to  say.  The event  of  Tim's  birth could be  explained  only  with 
reference to the future. 

\subsection{Popper's Pond}

\noindent
In  the non-mechanical mode of  time-travel,  involving 
advanced   interactions,  Tim's   `birth' corresponds exactly to Popper's 
pond paradox. If a stone is dropped into  a pond, ripples usually 
spread  outwards  (corresponding to a retarded wave). In the advanced 
case, the ripples  converge spontaneously and  throw  the stone  out  of  the 
pond. This sort  of  thing,  though possible  according to physical 
theory, is not  usually observed  unless one has filmed the sequence  and  plays 
the  film  backwards. But, says Popper,$^{15}$  `no  physicist would  mistake 
the end of the film for  its  beginning; for  the  creation  of  a  contracting  circular   wave 
followed  by  a  zone of  undisturbed  water  would  be (causally   considered)   miraculous.'   Popper's   own 
argument involved coherence: for constructive interference  of  primary  wavelets, 
to  produce  a  converging ripple,  by Huyghens' principle, one would 
need  coherence,  and  this  would be  practically  impossible  to 
arrange without `organization from the centre'. 

\begin{figure}[h]
\centerline{
\includegraphics
[width=18pc] {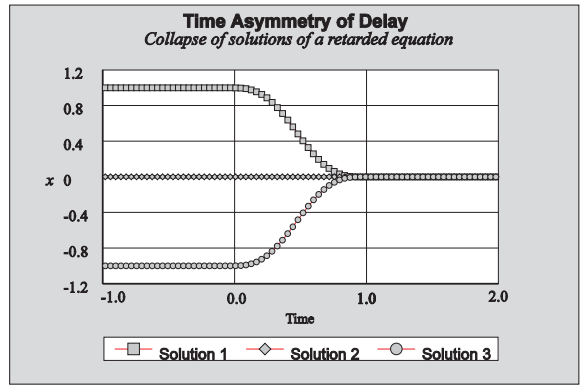}}
\caption{Three solutions of a retarded equation. The different 
past histories prescribed over [$-1$, 0] all result in the same future 
for $t \ge 1$. Retrodiction is hence impossible from future data 
prescribed over $t \ge 1$. Teleological explanations are impossible, 
with history-dependent evolution.}
\label{retarded}
\end{figure}

\begin{figure}[h]
\centerline{
\includegraphics
[width=18pc] {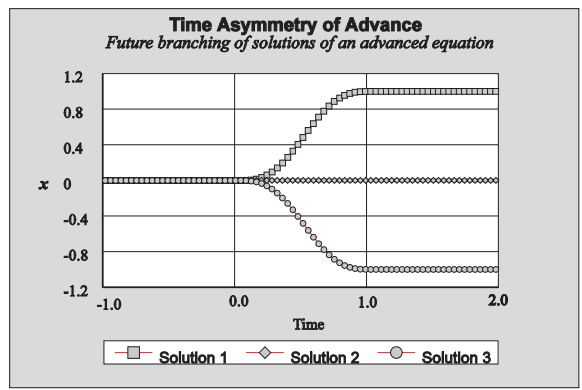}}
\caption{Three solutions of an advanced equation: different futures 
over [1,2] correspond to the same past for $t \le 0$. With anticipation 
past fails to decide the future, for one past may correspond to many 
futures, in this time reverse of Fig. \ref{retarded}. Hence, causal explanations 
are impossible with anticipatory evolution.
}
\label{advanced}
\end{figure}

\begin{figure}[h]
\centerline{
\includegraphics
[width=18pc] {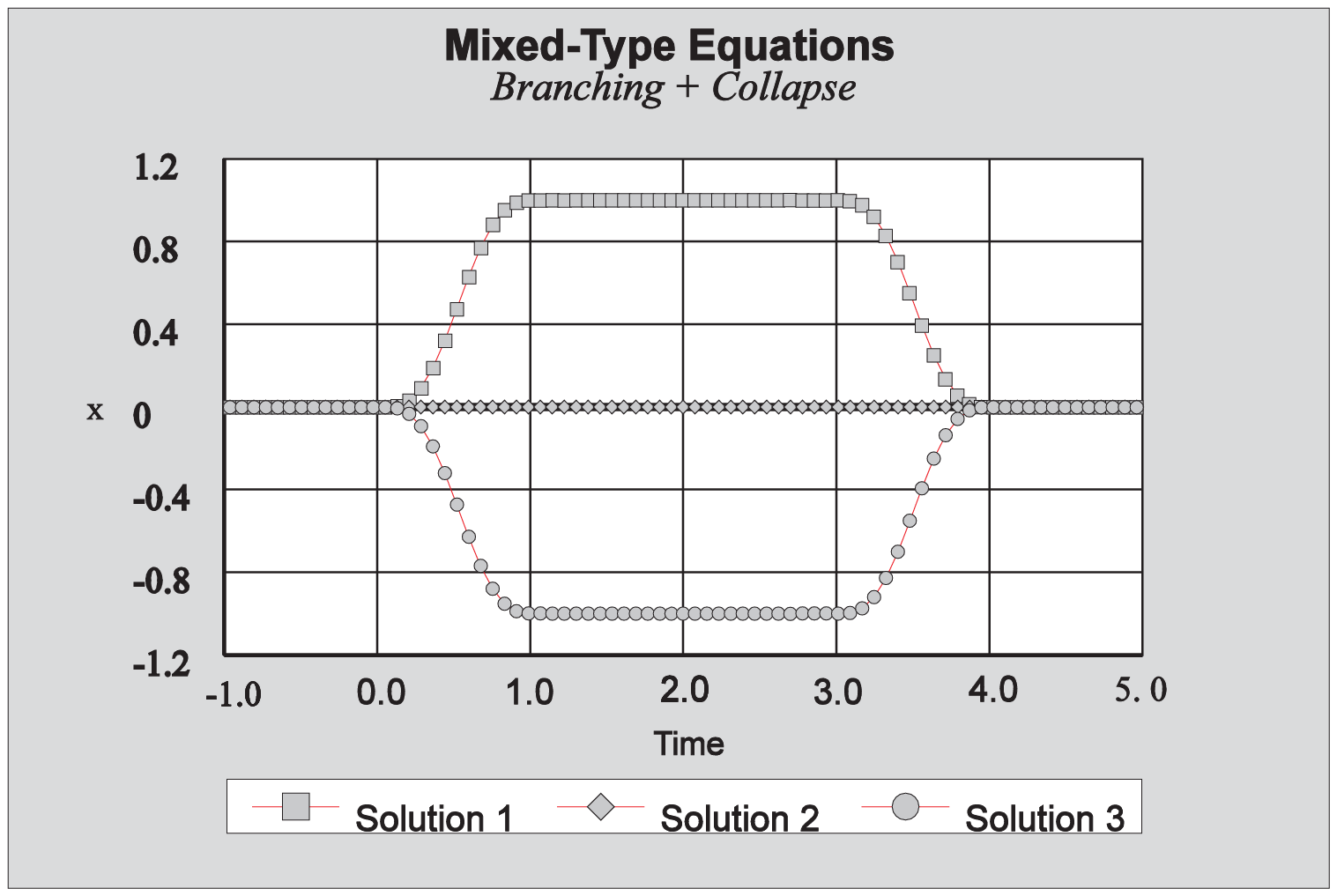}}
\caption{With a realistic mixture of history-dependence and a 
small amount of anticipation, the past still fails to decide the future. 
With this model,  all phenomena do not admit causal explanations, 
so that spontaneity really is possible. The existence of a small tilt 
is exactly the condition for time-travel of the second kind.}
\label{mixed}
\end{figure}

One can strengthen the first part of Popper's argument, by  giving  more 
general and stronger  arguments  which show   the  theoretical  impossibility  of   explaining 
anticipatory phenomena from the past. A causal explanation   of  anticipatory  phenomena  is   mathematically 
impossible for exactly the same reason that a teleological  explanation  of 
purely history  dependent  phenomena  is mathematically  impossible. 
Purely anticipatory phenomena  may be  explained only by reference 
to the future, just  as history  dependent phenomena may be explained  only  by 
reference to the past, for the reasons sketched in  Figs \ref{retarded}, \ref{advanced}, \ref{mixed}, 
reproduced from Ref. 7, where they are explained in more detail. (A quick exposition is also provided in the appendix to this note.)

The  pond paradox is now seen to arise from  Popper's  metaphysical  stipulation 
that all phenomena must  admit a causal  explanation, so that phenomena 
not admitting  a causal explanation cannot possibly occur. This position 
is  reminiscent  of the Stoics who  derived  \textit{heimarmene} (fate) 
from \textit{eiro} (string beads), so that the  evolution of  the  world  was  analogous 
to  moving  beads  on  a necklace; the slightest spontaneous swerve 
of the atoms (Epicurean  \textit{clinamen})  would  break  the  string:  `the 
cosmos  would  break  up  and  be  shattered\ldots if  some uncaused  movement  were 
to be introduced  into  it.$^{16}$ Perhaps it is necessary to restate 
that a  metaphysical stipulation  (`everything must have an antecedent 
cause'), as  used e.g.  by  Hippocrates, may not  be  used  to  decide 
admissible  phenomena. The existence or non-existence of the spontaneous 
can only be decided by observation. 

\subsection{The Empirical Evidence}

\noindent
The  absence of hordes of tourists from the future  is, 
therefore,  no  evidence  against time  travel  of  the second  kind.  It 
would be enough  if  we  occasionally observe some spontaneous events.

\subsection{The Mechanization of Spontaneity}

\noindent
A key feature of spontaneity in the above sense is that 
\textit{spontaneity cannot be mechanized}, i.e., though time travel 
may be possible, time \textit{machines} are not: only time travel of 
the second kind is possible. Popper's conclusions from his pond paradox 
only need to be toned down: while the existence of a causal explanation 
cannot very  well be a precondition for the occurrence of a  phenomenon, 
without a causal explanation one cannot systematically \textit{control}  the 
phenomenon, or \textit{arrange} for it to occur, or \textit{mechanically} 
reproduce it. The Wellsian time-machine incorporates in its construction 
the intuitive idea of `control' from the future. In the physics literature, the same idea was articulated in the context of the tachyonic anti-telephone: if Shakespeare used a tachyonic anti-telephone to dictate \textit{Hamlet} to Bacon then, Benford et al.$^{17}$ argued, while Bacon would have chronological priority, Shakespeare remained the author of \textit{Hamlet}---since Shakespeare was the one who had `control'. But, in a situation 
where interactions may propagate from future to past, it is not clear 
that control from the future is any more possible than control from 
the past, and Fig. \ref{mixed} sketches a counter-example: in some situations 
prescribing both past and future data may still be inadequate to determine 
a unique present. Similarly, the classical argument$^{18}$ to exorcise 
Maxwell's demon excludes only the mechanical form of  the demon, which 
could lead to a controllable, hence possibly unboundedly large, decrease 
of entropy. 

\begin{figure}[h]
\centerline{
\includegraphics
[width=18pc]{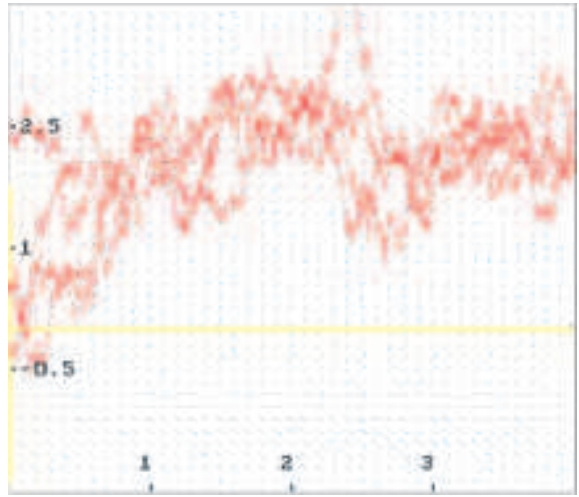}}
\caption{The figure shows some Brownian sample paths. The sample 
paths mix: the trajectories tend to `forget' their past and asymptotically 
become statistically independent of it, unlike the background vector 
field (dashed lines) which corresponds to solutions of the unperturbed 
ordinary differential equation. `Mixing' due to chance is believed 
to produce entropy rather than order (= negentropy).}
\label{brownian}
\end{figure}

\subsection{Spontaneity and Chance}

\noindent
The relevance of Maxwell's demon is the following. Spontaneity, 
in the sense proposed above, differs from the notion of `chance' in 
the sense of probabilistic (`stochastic') evolution, such as that 
of a stochastic process, where the probabilities of future states 
can be computed once the past states are known. (The meaning we have 
assigned to `chance' is related to contemporary customary usage amongst 
physicists: for the last hundred years or so, it has been argued that 
probabilistic evolution accounts for entropy increase within classical 
reversible dynamics.) Mathematically, the difference is that evolution 
involving such `chance' may be modeled by stochastic differential 
equations (Fig.\ref{brownian}), categorically distinct from the mixed-type functional differential 
equations which model evolution involving a tilt in the arrow of 
time. In physical terms, a key difference is that chance corresponds 
to `mixing' while spontaneity, in the above sense,  corresponds to 
`sorting'. 

That is, if the implicit assumption of some kind of `mixing' 
or quasi-ergodicity is acceptable as a characteristic feature of `chance' 
in the sense of  probabilistic evolution, one  might  say that `chance' 
results in an  increase  of entropy, whereas spontaneity in the above 
sense (e.g. a converging  ripple) clearly corresponds to a  reduction 
in entropy, or to the creation of order. So, time travel of the second 
kind actually corresponds to spontaneous order creation.  

Though general relativistic statistical mechanics (and the statistical 
basis of the stress-energy continuum) is problematic,  the apparently 
necessary association of exotic matter (hence negative energies) with 
wormhole spacetimes  suggests, in the  standard  MTW$^{19}$ approach 
to relativistic thermodynamics, that, one should expect a similar 
association of entropy  reduction  with  time travel in  the  case  of 
wormholes. 

It is natural to conjecture that any macrophysical manifestation of 
spontaneity would involve, in an essential way, the one thing that 
has remarkably resisted mechanization: life. Specifically, I expect 
that a systematic microphysical tilt in the arrow of time would show 
up in the structure and dynamics of biological macromolecules. At 
present, solutions of the many-particle equations of motion with a 
microphysical tilt in the arrow of time are still being simulated, 
and compared with solutions of a stochastically perturbed form of 
the classical equations, and only preliminary results are available,$^{20, ~ 21}$ 
so that it would not be in order to make a more definite statement. 
However, some general arguments connecting spontaneity in the above 
sense with `human freedom' in the mundane sense of Fig. \ref{mundane} may be found 
in Ref. 7.

At the microphysical level, spontaneity as a necessary correlate of non-locality is especially interesting in the context (Ref. 7) of the structured-time interpretation of quantum mechanics. 

\section{CONCLUSIONS}

\noindent
Time travel conflicts not with choice but with `causality': 
if two-way interaction with the future is permitted, one can no longer 
hang on to `causality' in the sense of demanding explanations exclusively 
from the past. Interactions propagating from future to present imply 
the occurrence of events that are causally inexplicable. Under the 
circumstances of  time travel, one must allow for the reality of such  spontaneous 
events, which differ from `chance' events in creating order instead 
of destroying it. The mechanization of spontaneity, however, is impossible, 
so that time travel can only be of the second kind.  

\appendix

\textit{A causal explanation of anticipatory phenomena is mathematically impossible:} following the referee's suggestion to keep the paper self-contained, we reproduce here from Ref. 7, some  mathematical details of the argument.

First, let us see why \textit{a teleological explanation of history-dependent phenomena is mathematically impossible}. Fig. \ref{retarded} shows three solutions of the retarded FDE

\begin{equation}
x ' (t) = b(t) x(t-1) ,
\label{eqret}
\end{equation}

\noindent where $b$ is a continuous function which vanishes outside [0,~1], and satisfies

\begin{equation}
\int b(t)~dt = -1 .
\label{b}
\end{equation}

\noindent For example, 

\begin{equation}
b(t) = \left  \{  
\begin{array}{r @{\quad : \quad} l}
0 & t \le 0 \\
-1 + \cos 2 \pi t & 0 \le t \le 1 . \\
0 & t \ge 1
\end{array}
\right .
\end{equation}

For $ t \le 0 $, the FDE \eqrefa{eqret} reduces to the ODE $x' (t) = 0$ , so that, for $t \le 0$,  $x(t) = k$ for some constant $k$ (= $x(0)$).

Now,  for $t \in [0,~1]$, 
\begin{eqnarray}
x(t) &=&  x(0) + \int_0^t x'(s) ds \nonumber \\
\; &=& x(0) + \int_0^t b(s) x(s-1) ds \nonumber \\
&=& x(0) + x(0) \int_0^t b(s) ds ,
\end{eqnarray}

\noindent since $x(s-1) \equiv k = x(0)$ for $s \in [0, 1]$. Hence,  using \eqrefa{b}, $x(1) = 0$, no matter what $k$ was. However, since $b(t)=0$ for $t \ge 1$, the FDE \eqrefa{eqret} again reduces to the ODE $x' (t) = 0$, for $t \ge 1$, so that $x(1) = 0$ implies $x(t) = 0 $ for all $t \ge 1$. Hence, the past of a system governed by \eqrefa{eqret} cannot be retrodicted from a knowledge of the entire future; for if the future data (i.e., values of the function for \textit{all} future times $t \ge 1$) are prescribed using a function $\phi$ that is different from 0 on $[1,~ \infty ]$, then \eqrefa{eqret} admits no backward solutions for $t \le 1$. If, on the other hand,  $\phi \equiv 0 $ on  $[1,~ \infty ]$, then there are an infinity of distinct backward solutions. In either case, knowledge of the entire future furnishes no information about the past. 

The actual solutions shown in the graph were obtained numerically, using the \textsc{retard} package of Hairer et al.$^{22}$ 

In the advanced case, as suggested by Fig. \ref{advanced}, the argument is the time-symmetric counterpart of the above argument. In this case, the equation solved was the analogous advanced FDE
\begin{equation}
x'(t) = b(t) x(t+1) ,
\label{eqadvanced}
\end{equation}

\noindent where the function $b$ has the same properties as before, except that

\begin{equation}
\int b(t) dt = 1 .
\label {b1}
\end {equation}

For example, 

\begin{equation}
b(t) = \left  \{  
\begin{array}{r @{\quad : \quad} l}
0 & t \le 0 \\
1 - \cos 2 \pi t & 0 \le t \le 1 . \\
0 & t \ge 1
\end{array}
\right .
\end{equation}

The reasoning proceeds in an entirely analogous manner. For $ t \ge 1 $, the FDE \eqrefa{eqadvanced} reduces to the ODE $x' (t) = 0$, so that, for $t \ge 1$,  $x(t) = k$ for some constant $k$ (= $x(1)$).

Now,  for $t \in [0,~1]$, 
\begin{eqnarray}
x(t) &=&  x(1) - \int_t^1 x'(s) ds \nonumber \\
\; &=& x(1) - \int_t^1 b(s) x(s+1) ds \nonumber \\
&=& x(1) - x(1) \int_t^1 b(s) ds  ,
\end{eqnarray}

\noindent since $x(s+1) \equiv k = x(1)$ for $s \in [0, 1]$. Hence,  using \eqrefa{b1}, $x(0) = 0$, no matter what $k$ was. However, since $b(t)=0$ for $t \le 0$, the FDE \eqrefa{eqadvanced} again reduces to the ODE $x' (t) = 0$, for $t \le 0$, so that $x(0) = 0$ implies $x(t) = 0 $ for all $t \le 1$. Hence, the future of a system governed by \eqrefa{eqadvanced} cannot be predicted from a knowledge of the entire past; for if the past data (i.e., values of the function for \textit{all} past times $t \le 0$) are prescribed using a function $\phi$ that is different from 0 on $[- \infty, ~0 ]$, then \eqrefa{eqadvanced} admits no forward solutions. If, on the other hand,  $\phi \equiv 0 $ on  $[- \infty, 0 ]$, then there are an infinity of distinct forward solutions. In either case, precise knowledge of the entire past furnishes no information about the future. The actual numerical solutions shown were obtained by a time-symmetric modification of the \textsc{retard} package.

Fig. \ref{mixed} shows solutions of the mixed-type equation

\begin{equation}
x'(t) = a(t) x(t-1) + b(t) x(t+1) ,
\end{equation}

\noindent where $b$ has the same properties as in \eqrefa{b1}, and the continuous function $a$ now has support on the interval $[2, ~3]$, and satisfies

\begin{equation}
\int_2^3 a(t) dt = -1  .
\end{equation}

\noindent The solutions may be obtained by combining the reasoning used in the preceding two cases. 

Physically, retarded FDE arise as the equations of motion of charged particles, using the Heaviside-Lorentz force law, and assuming fully retarded Lienard-Wiechert potentials.$^{7, ~20, ~21}$
Mixed-type equations arise as the equations of motion of charged particles in the case where most electromagnetic radiation is retarded, but some of it may be advanced, i.e., we use a convex combination of retarded and advanced Lienard-Wiechert potentials. This possibility has often been excluded on metaphysical grounds, without studying the immediate empirical consequence (of spontaneity), here and now, of this assumption.   

Finally, in the case of Fig. \ref{brownian} the equation solved was a stochastic differential equation of the type

\begin{equation}
dX_t = a(t, ~X_t) dt + b(t,~ X_t) dw(t) ,
\end{equation}

\noindent where $w(t)$ is the standard Brownian motion (Wiener process). The background vector field relates to the deterministic part of this equation, obtained using only the drift function $a(t, ~X_t)$ and setting the dispersion function $b(t, ~X_t)$ to zero.
The sample paths shown in the figure were obtained using this author's package \textsc{stochode} for the solution of stochastic differential equations (SDE's) driven by Brownian or L\'evy motion. 

Given the vast difference between the mathematical theory underlying SDE's (`chance') and that underlying mixed-type FDE's (`spontaneity') it is surprising why it should be hard to discriminate between the physical consequences of the two.  In the case of SDE's (`chance') the future is \textit{epistemically} uncertain since (a) the past is uncertain, and (b) the relation of past to future is probabilistic rather than deterministic.  In the case of mixed-type FDE's (`spontaneity'), the future is \textit{ontically} uncertain, regardless of knowledge of the past, because past does not entirely determine the future   

\section*{REFERENCES}

\noindent
1. K. S. Thorne,  in:  R. J. Gleiser, C.N. Kozameh  and  O. M.  Moreschi 
(eds),  \textit{General   Relativity  and  Gravitation 1992} (Institute  of  Physics,  Bristol, 
1993) pp. 294--315; K.S.  Thorne,  \textit{Black  Holes   and  Time  Warps: 
Einstein's  Outrageous Legacy} (Norton, New  York, 1994); M.S. Morris 
and K.S. Thorne, \textit{Amer. J. Phys.}  \textbf{56} 395--412 (1988) ;  M.S.  Morris,   K.S.  Thorne,   and 
U. Yurtserver, \textit{Phys. Rev. Lett.}  \textbf{61} 1446 (1988) ;   S.  -W.  Kim  and  K.S.  Thorne,  \textit{Phys. 
Rev. D} \textbf{44} 4735--37 (1991).

\noindent
2. R.  H. Price,  \textit{Amer. J. Phys}.  \textbf{61} 
216--7 (1993).

\noindent
3. J.  R. Gott  III,  \textit{Phys. Rev. Lett.} \textbf{66}  1126--29  (1991).

\noindent
4. S. W. Hawking, \textit{Phys. Rev.  D} \textbf{46}  
603--11 (1992).

\noindent
5. Instantaneous transfer of information at superluminal 
speeds naturally does not contradict the theory of relativity,  which 
only requires that the speed of light should be constant in any local 
Lorentz frame. The  usual  assertion of a conflict between the two 
implicitly assumes `causality' which is not quite applicable under 
the circumstances.

\noindent
6.  Contrary  to the claims made in the current literature, 
the Wheeler-Feynman absorber theory may  not  be  used for this purpose 
since it is internally inconsistent, while the Hoyle-Narlikar absorber 
theory may not be used  since it is externally inconsistent. See, 
C. K. Raju, \textit{J. Phys. A: Math. Gen.} \textbf{13} 3303--17 (1980).

\noindent
7.  C. K. Raju, \textit{Time: Towards a Consistent Theory} 
(Kluwer Academic, Dordrecht, 1994). Since retarded  interactions dominate, 
one may still distinguish between forward and backward directions 
of time.

\noindent
8. For this purpose, various subtle distinctions such 
as the distinction between Wellsian and G\"{o}delian forms of time 
travel  do not seem critical. See, John Earman, ``Recent work on time 
travel." In: Steven  F. Savitt (ed.), \textit{Time's Arrows Today}  (Cambridge 
University Press, Cambridge, 1995).

\noindent
9. The presentation is adapted from D. Lewis, in: R. L. Poidevin and M. Macbeath  (eds), \textit{The Philosophy of Time} (Oxford University Press, Oxford, 1993), pp. 134--46.

\noindent
10. J. F. Woodward, \textit{Found. Phys. Lett.} \textbf{8} 
1--39 (1995), p. 2.

\noindent
11. In the sense of temporal logic, see, e.g. N. Rescher 
and A. Urquhart, \textit{Temporal Logic} (Springer, Wien, 1973);  W. 
H. Newton Smith, \textit{The Structure of Time} (Routledge and Kegan 
Paul, London, 1980).

\noindent
12.  K.  R.  Popper,  \textit{The  Open Universe: an Argument 
for Indeterminism}, Post  Script  to  \textit{The  Logic  of  Scientific 
Discovery}, Vol. 2 (Hutchinson, London, 1982), p. 58n--59n.

\noindent
13.  S.  W.  Hawking  and  G.  F. R. Ellis, \textit{The Large  Scale  Structure  of  Spacetime} (Cambridge  University  Press, 
Cambridge, 1973), p. 183.

\noindent
14. A number of differing `non-standard' views have, 
however, been documented by P. J. Nahin, \textit{Time Machines: Time Travel 
in Physics, Metaphysics, and Science Fiction} (American Institute 
of Physics, New York, 1993).

\noindent
15. See K. R. Popper cited in Note 12 above. The original 
discussion was in K. R. Popper, \textit{Nature} \textbf{177} 538 (1956); 
\textbf{178} 382  (1956); \textbf{179} 297 (1957); \textbf{181} 402 (1958).

\noindent
16.  Alexander  Aphrodisiensis,  \textit{De fato}, 192, 
6, cited in S. Sambursky, \textit{Physics of the  Stoics} (Routledge  and  Kegan 
Paul, London, 1987), p. 57.

\noindent
17. G. A. Benford, D. L. Book, and W. A. Newcomb, \textit{Phys. Rev.} D \textbf{2} 263--65 (1970).

\noindent
18. L. Szilard, \textit{Z. Phys}., \textbf{53} 840  (1929); 
L. Brillouin, \textit{J. Appl. Phys.}, \textbf{22} 334 (1951).

\noindent
19 C.W. Misner, K. S. Thorne, and J.A. Wheeler, \textit{Gravitation}, 
(W. H. Freeman, San Francisco, 1978).

\noindent
20. C. K. Raju, `Simulating a Tilt in the Arrow of Time: 
Preliminary Results.' Paper presented at a Seminar on `Some Aspects 
of Theoretical Physics', Indian Statistical Institute, Calcutta, 14--15 
May 1996.

\noindent
21. C. K. Raju, `The Electrodynamic Two-Body Problem and the Origin of Quantum Mechanics,' \textit {Found. Phys.} \textbf{34} 937--62 (2004).

\noindent
22. E. Hairer, S. P. Norsett, and G. Wanner, \textit{Solving Ordinary Differential Equations} (Springer Series in Computational Mathematics, Vol. 8) (Springer, Berlin, 1987).

\end{document}